\documentclass[twocolumn,superscriptaddress,amssymb,amsmath,nobibnotes,aps,prl]{revtex4-2}

\usepackage{color}
\usepackage{comment}
\usepackage{graphicx}
\usepackage{tensor}

\usepackage{braket}

\newcommand{\dd}{\mathop{\mathrm{d}\!}{}}

\newcommand{\ri}{r_{\text{int}}}

\makeatletter
\def\@affil@script#1#2#3#4{%
 \@ifnum{#1=\z@}{}{%
  \begingroup
   \frontmatter@affiliationfont
   \@ifnum{\c@affil<\affil@cutoff}{}{%
    \def\@thefnmark{#1}\@makefnmark
   }%
   \ignorespaces#3%
   \@if@empty{#4}{}{\frontmatter@footnote{#4}}%
  \endgroup
 }%
}%
\makeatother

\begin{document}

\title{Quantum Field Theory on compact stars near the Buchdahl limit}

\author{Ignacio A. Reyes}
\affiliation{Institute for Theoretical Physics and\; }
\author{Giovanni Maria Tomaselli}
\affiliation{GRAPPA,\\ University of Amsterdam,  Amsterdam, 1098 XH, The Netherlands}


\begin{abstract}
Very compact stars seem to be forbidden in General Relativity. While Buchdahl's theorem sets an upper bound on compactness, further no-go results rely on the existence of two light rings, the inner of which has been associated to gravitational instabilities. However, little is known about the role of quantum fields in these strong gravity regimes. Here, we consider the particularly simple model of a constant density star and we work in the probe approximation where the backreaction is ignored. We show that the trapping of modes inside the star leads the renormalized stress tensor of Conformal Field Theories to diverge faster than the classical source in the Buchdahl limit. This leads to the violation of the Null Energy Condition around the inner light ring. The backreaction of quantum fields in this regime therefore cannot be ignored. This happens as the star's surface approaches the Buchdahl radius $9GM/4$ rather than the Schwarzschild radius. The results are independent of the details of the interactions, but contain an ambiguity associated to the renormalization scheme.
\end{abstract}

\maketitle

\section{Compact relativistic stars}

The General Relativistic prediction of the existence of compact objects, such as white dwarfs and neutron stars, has been confirmed by many observations. Their macroscopic properties follow from the Tolman-Oppenheimer-Volkoff equation. However, quantum theory is essential in understanding the physics of these stars, as it provides the ultimate reason for their existence, namely, Fermi's exclusion principle.

The question regarding the maximum mass of such compact object is crucial: it is the main criterion used to discriminate between what we suspect is a neutron star or a black hole. Well-known upper limits were set by Chandrasekhar\,\cite{Chandrasekhar:1931ih} and Rhoades-Ruffini\,\cite{Rhoades:1974fn}. A more generic result, which is independent of the equation of state of the matter, was established by Buchdahl\,\cite{Buchdahl:1959zz} and gives an upper bound on compactness in GR. Consider an isotropic perfect fluid star, with stress tensor
\begin{align}\label{Tmunufluid}
\tensor{T}{^\mu_\nu}=\text{diag}(-\rho,p,p,p)\,,
\end{align}
on a static, spherically symmetric metric
\begin{align}\label{ds2}
   ds^2 &=-f(r)\dd t^2+h(r)\dd r^2+r^2 (\dd \theta^2+\sin^2\theta \dd \phi^2)\,.
\end{align}

Assuming in addition that $\rho>0$, $\partial_r\rho\le0$ and that Einstein's equations hold, the requirement that the metric is everywhere regular leads to
\begin{align}\label{BL}
R\ge 9GM/4\,,
\end{align}
where $R$ is the radius of the star and $M$ is its mass. The saturation of the bound is known as the Buchdahl limit. Notice one can also formulate this bound in a coordinate-independent way.

A particularly simple solution that manifestly saturates Buchdahl's second assumption is the constant-density star or ``Schwarzschild interior metric.'' These configurations have uniform density $\rho=3M/4\pi R^3$ throughout the star, and as is well known they can saturate the Buchdahl limit \eqref{BL}. Although they are unrealistic models of an astrophysical object, they are the standard example when studying the TOV equations. 

The metric for this equation of state takes the form (\ref{ds2}), with
\begin{align}
    f(r)&=\left( \frac{3}{2}\sqrt{1-\frac{2GM}{R}}-\frac{1}{2}\sqrt{1-\frac{2GMr^2}{R^3}} \right)^{2}\,, \label{f} \\
    h(r)&=\left( 1-\frac{2GMr^2}{R^3}  \right)^{-1}\,, \label{h}
\end{align}
and is matched to the usual exterior Schwarzschild vacuum solution at the sphere's surface. 

The simplicity of these solutions make them an excellent setup to test Quantum Field Theory (QFT) in the strong gravity regime. A final motivation to consider this metric is that it is conformally (Weyl) flat. In fact, the uniform density metric above is the unique solution to Einstein's equations coupled to a static perfect fluid that is conformally flat\,\cite{1971AmJPh..39..158B,Raychaudhuri1979ConformalFA}. This will allow us to obtain explicit analytic results. We will thus work with this spacetime, and comment about the generality of our results later on.

\section{Wave equation}

\begin{figure}
      \centering
      \includegraphics[width=1\columnwidth,trim={0 6pt 0pt 0}]{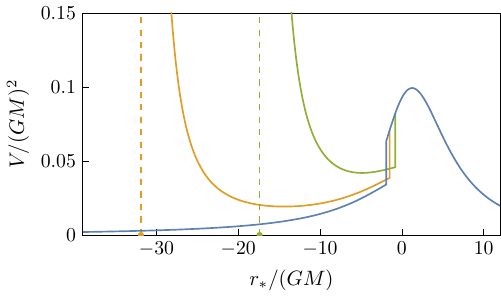}
      \caption{Potential (for $\ell=1$) given in (\ref{eqn:potential}), for $R/(GM)=9/4$ (blue), $2.3$ (orange) and $2.4$ (green). The dashed lines mark the the values of $r_*$ corresponding to the center of the star in the two latter cases. A discontinuous jump at the star's surface matches it to the exterior vacuum Schwarzschild solution.}
      \label{fig:potential}
\end{figure}

In order to understand the behavior of quantum fields in this spacetime, let us begin by first considering the propagation of classical waves in it. The wave equation for the uniform density background was first discussed by Chandrasekhar and Ferrari\,\cite{1991RSPSA.434..449C}. For simplicity we take a massless scalar $\Phi$ using the usual decomposition
\begin{align}
\Phi=\sum f_{\omega\ell m}\,,\qquad f_{\omega\ell m}(x)=\frac{u(r)}rY_{\ell m}(\theta,\phi)e^{-i\omega t}\,.
\end{align}
The wave equation $\Box\Phi=0$ can be recast in a Schrödinger-like form:
\begin{align}
-\partial_{r_*}^2u+V(r_*)u=\omega^2u\,,
\end{align}
where we defined the tortoise coordinate $r_*$ via
\begin{align}
\frac{\dd r_*}{\dd r}=\sqrt{h(r)/f(r)} \,.
\end{align}
The potential $V(r_*)$ takes the form
\begin{align}
V=\frac1r\partial_{r_*}^2r+\frac{\ell(\ell+1)}{r^2}f
\label{eqn:potential}
\end{align}
and is plotted in Fig.\:\ref{fig:potential} for $\ell=1$ and various values of $R/(GM)$. The potential at $r>R$ corresponds to the Schwarzschild vacuum metric, and vanishes at infinity. It connects to the interior of the star with a discontinuous step.

As we can see from Fig.\:\ref{fig:potential}, when $R>9GM/4$, the tortoise coordinate has a finite minimum possible value (dashed lines) corresponding to the center of the star, because the factor $h/f$ is always regular around $r=0$. Moreover, $V(r_*)$ reaches a local minimum greater than zero and then increases towards the surface.

When $R\to9GM/4$, however, one has $h/f\sim r^{-2}$ and therefore the domain of $r_*$ becomes infinite on both sides, while $V(r_*)$ vanishes at the center of the star. This closely resembles the situation for black holes, but in that case it is the horizon that is mapped to $r_*\to-\infty$. If the star is sufficiently close to the Buchdahl limit, then the field modes can thus be \emph{trapped} inside it, leading to a spectrum of quasibound states whose magnitudes are amplified close to the origin.

These properties of the effective potential, together with the behavior of the tortoise coordinate, suggests that upon quantization the renormalized stress tensor can become important in the Buchdahl limit. The rest of our analysis will be done in a more generic way that depends less on the specific theory considered.

\section{QFT in curved spacetime}

QFT in curved spacetime has seen significant progress in the last half century. In the semiclassical approximation, gravity is still treated classically and one considers some quantum fields as another dynamical source to Einstein's equations,
\begin{align}\label{semi-cl-einstein}
\mathcal{R}_{\mu\nu}-\frac12g_{\mu\nu}\mathcal{R}=8\pi G\bigl(T_{\mu\nu}+\braket{\hat T_{\mu\nu}}\bigr)\ .
\end{align}
We shall denote by $\hat T_{\mu\nu}$ the operator of the QFT to distinguish it from the classical source \eqref{Tmunufluid}. 

However, most work in this field has focused on either cosmology or black holes. Here, we will study the role it plays for astrophysical compact stars. The question we will address in this work is whether there exists some \textit{generic} feature of QFT, independent of the details of the nuclear interactions and the quantum states involved, that becomes important for very compact stars, in the regime of strong gravitational fields. We will show that there is indeed such an effect.

We shall focus on the effects of conformally coupled fields where the computation is easier, taking it as a toy model for more generic scenarios. We work in $3+1$ dimensions, but the generalization to even higher dimensions is straightforward. The nonconformal case will be treated elsewhere.

As is well known, conformally coupled classical matter has a vanishing trace of its stress tensor. However, its quantum counterpart develops a \textit{trace anomaly}. In $3+1$ dimensions, the vacuum expectation value of the trace of the renormalized stress tensor for quantum fields propagating in a curved spacetime is 
 \begin{align}\label{Tmumu}
    \braket{ \tensor{\hat T}{^\mu_\mu} } = \frac{1}{(4\pi)^2}\left[ c \mathcal{F} - a \mathcal{G} - d \Box \mathcal{R}  \right]\,,
\end{align}
where $\mathcal{R}$ is the Ricci scalar, $\mathcal{F}$ is the square of the Weyl tensor and $\mathcal{G}$ is the Gauss-Bonnet invariant. Amongst the three real coefficients, $c>0$ and $a>0$ are well understood and characterize the particular theory in question. On the other hand, $d$ is not determined by the bare Lagrangian as it depends on the renormalization scheme, and is closely related to the quadratic corrections to the gravity action as we review below. As such, it should be fixed by experiments. For now we will leave $d$ as a fixed but undetermined constant and proceed with the calculation. 

If additionally the metric is conformally flat---as is the case for the constant-density star---then all components of the renormalized stress tensor are fixed\,\cite{Brown:1977sj}:
\begin{equation}
\begin{split}
	&\braket{\hat T^{\mu\nu}}=\\
    &-\frac{a}{(4\pi)^2}\biggl[ g^{\mu\nu} \biggl( \frac{ \mathcal{R} ^2}{2}-\mathcal{R}_{\alpha\beta} \mathcal{R}^{\alpha\beta} \biggr) + 2 \mathcal{R}^{\mu\lambda}\tensor{\mathcal{R}}{^\nu_\lambda} -\frac{4}{3}\mathcal{R} \mathcal{R}^{\mu\nu} \biggr]\\
    &+\frac{d}{(4\pi)^2} \biggl[ \frac{1}{12} g^{\mu\nu} (\mathcal{R}^2-4\tensor{\mathcal{R}}{^{,\lambda}_{;\lambda}})-\frac{1}{3}(\mathcal{R} \mathcal{R}^{\mu\nu}-\mathcal{R}^{,\mu;\nu}) \biggr] .\label{Tmunuq}
\end{split}
\end{equation}
The quantum state chosen for \eqref{Tmunuq} is the vacuum, but this will not play an important role. A state with finite temperature or with a fixed number of particles would only add an extra contribution (independent of the curvature) that remains finite in the Buchdahl limit. The vacuum stress tensor for the interior of the uniform density star is therefore given by \eqref{Tmunuq}. We now proceed to evaluate it and examine its properties.

\section{Quantum fields in the Buchdahl limit}

In this section, we describe the main features of the quantum stress tensor \eqref{Tmunuq} evaluated on the Schwarzschild interior metric. In particular, we wish to understand its behavior as we approach the Buchdahl limit
\begin{align}\label{BL1}
R=(9/4+\epsilon)GM\,,\qquad\epsilon \to  0\,.
\end{align}
We will report the results to leading orders in $\epsilon$.

The Buchdahl limit \eqref{BL1} is a finite distance above the black hole compactness corresponding to $R=2GM$. Nevertheless, this regime is no less extreme: the Ricci scalar $\mathcal{R}$ of the background metric at the center diverges in this limit as
\begin{align}\label{Ricci0}
\mathcal{R}(0) =-\frac{3}{R^2\epsilon}+\mathcal{O}(1)\,.
\end{align}
Correspondingly, the central density and pressure of the classical uniform density star solution behave as
\begin{align}
\rho(0)&=\frac{1}{3\pi GR^2}+\mathcal{O}(\epsilon)\,, \label{rhoclass0} \\
p(0)&=\frac{1}{8\pi GR^2 \epsilon} + \mathcal{O}(1)\,. \label{pclass0}
\end{align}

\begin{figure}
      \centering
      \includegraphics[width=1\columnwidth,trim={0 6pt 0pt 0}]{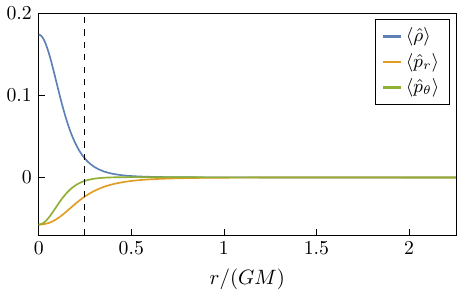}
      \caption{ Radial profile of the three components of $\langle \tensor{\hat T}{^\mu_\nu} \rangle$, for $a=d=1/360$, $\epsilon=0.003$, in units where $GM=1$. The location of the inner light ring is depicted by the dashed line.}
      \label{fig:plot}
\end{figure}

Let us contrast this behavior with its quantum counterpart \eqref{Tmunuq}. For generic $\epsilon$, this takes the form
\begin{align}\label{Tmunuqstar}
\langle \tensor{\hat T}{^\mu_\nu}\rangle = \text{diag}(-\langle \hat \rho \rangle,\langle \hat p_r \rangle,\langle \hat p_{\theta} \rangle,\langle \hat p_{\theta} \rangle)\,,
\end{align}
with $\langle \hat p_r \rangle\neq \langle \hat p_\theta \rangle$. The radial dependence of the components are illustrated in Fig.\:\ref{fig:plot}. 
In the limit $\epsilon\to 0$, their central values scale as
\begin{align}
\hspace{-.4cm}\langle \hat \rho(0) \rangle&= \frac{9d}{(8\pi R^2\epsilon)^2} + \frac{d}{6(\pi R^2)^2\epsilon}+ \mathcal{O}(1)\,,  \label{rhoq0} \\
\hspace{-.4cm}\langle \hat p_r(0) \rangle=\langle \hat p_\theta(0) \rangle&=-  \frac{3d}{(8\pi R^2\epsilon)^2} + \frac{2a-d}{(3\pi R^2)^2\epsilon}+ \mathcal{O}(1)\,. \label{pq0}
\end{align}
We emphasize that the pressures match only at the center, and not elsewhere. Moreover, notice that the leading order of $\langle \hat \rho\rangle$ and $\langle \hat p\rangle$ have opposite signs. There is no contribution from $c$ because the Weyl tensor vanishes.

By comparing \eqref{rhoclass0} and \eqref{pclass0} with \eqref{rhoq0} and \eqref{pq0}, we see that the components of the renormalized stress tensor scale with higher powers of $\epsilon$ than the classical contributions, and therefore cannot be ignored in the Buchdahl limit. Furthermore, notice that the leading divergence of the quantum terms depends only on $d$: in this regime the quantum effects are dominated by the scheme-dependent terms proportional to $d$, and not by $c$ or $a$.

The backreaction of quantum effects cannot be neglected if they become of the same order as the classical ones. By comparing the classical and quantum central pressures \eqref{pclass0} and \eqref{pq0}, this crossover happens at $\epsilon \sim |d| \left(\ell_P/R \right)^2$, which corresponds to a pressure $p\sim |d|^{-1}\ell_P^{-4}$ and central curvature $\mathcal{R}\sim -|d|^{-1}\ell_P^{-2}$, where $\ell_P$ is the Planck length. If $d\ll 1$, then this corresponds to sub-Planckian lengths and therefore we cannot trust our semiclassical analysis. Instead, if $d\gg 1$, then the QFT effects cannot be neglected in this regime. For this specific equation of state, a different, earlier crossover is found if the energy densities are compared instead. However, the impact of the constant energy density on the metric is negligible compared to that of the diverging central pressure, in the Buchdahl limit.

We will not address the problem of full backreaction in this work. Nevertheless, some general features of a linearized approximation provide useful insight. Consider the trace of the semiclassical equations \eqref{semi-cl-einstein},
\begin{align}\label{semi-cl-trace}
-\mathcal{R}=8\pi G\, ( -\rho+3p-\langle\hat{\rho}\rangle + \langle \hat{p}_r\rangle + 2\langle \hat{p}_\theta \rangle )\,,
\end{align}
evaluated at the origin, as we approach the Buchdahl limit. In the absence of the quantum corrections, the right-hand side of \eqref{semi-cl-trace} diverges as $\epsilon^{-1}$ as shown in \eqref{Ricci0}. However, as we see from \eqref{rhoq0} and \eqref{pq0}, the quantum contributions of the last three terms scale as $-d \epsilon^{-2}$.  

If $d<0$, then the quantum terms on the right side of \eqref{semi-cl-trace} grow without bound with the \textit{same} sign as the classical ones. This suggests a runaway: as the curvature increases, so do quantum effects, which increase the curvature further and so on. Conversely, if $d>0$, then the quantum contributions to the trace have the opposite sign, which \textit{decreases} the curvature. This suggests the possible existence of a backreacted solution, but only for $d>0$. Such an equilibrium would require a small but finite $\epsilon$ of the order discussed above, so the surface of such an object would lie very close the Buchdahl radius, and far from the Schwarzschild radius. 

\section{Role of the light ring}

Light rings (photon spheres) play a key role in our analysis. These are defined as regions where null geodesics form circles, and they always come in pairs due to topological arguments\,\cite{Cunha:2017qtt}. For $9GM/4<R\leq 3GM$, the above metric develops two light rings located at:
\begin{align}\label{LR}
 r_{\text{ext}}=3GM\ \ ,\ \    r_{\text{int}}&=\frac{1}{3}\sqrt{ \frac{R^3}{GM} \frac{4R-9GM}{R-2G M}  }\,.
\end{align}
The outer ring $r_{\text{ext}}$, also present for black holes, corresponds to the usual photon sphere outside the surface of the star and is unstable: photons crossing it either escape to infinity or spiral inwards. It has been probed by recent observations \cite{LIGOScientific:2016lio,EventHorizonTelescope:2019dse,Cardoso:2019rvt}.

The inner ring $\ri$ lies in the interior and is a \textit{stable} attractor of null geodesics, meaning that massless fields remain trapped around it. Notice that it shrinks to the origin in the Buchdahl limit. As illustrated in Fig.\:\ref{fig:plot}, the magnitude of the quantum stress tensor \eqref{Tmunuq} is maximum at the center and falls steeply around the inner light ring. Indeed, in the Buchdahl limit the inner light ring sets the location at which the field values have dropped roughly by one order of magnitude, i.e.
\begin{align}\label{rhofalloff}
\frac{\langle \hat \rho(\ri)\rangle}{\langle \hat \rho(0)\rangle}\sim 0.1
\end{align}
and similarly for the pressures. This shows that the region inside the inner photon sphere is where the quantum fields have most support, which is the quantum analog to the classical trapping of modes discussed above using the wave equation. The crossover when the classical and quantum pressures become comparable corresponds to an inner light ring of radius $r\sim \sqrt{|d|}\,\ell_P$. 

The inner photon sphere plays yet another important role: it is the location where the Null Energy Condition (NEC) is violated. Given a null vector $k^\mu$, one defines an operator by contracting the (total) stress tensor with it 
\begin{align}\label{NEC}
    \text{NEC}=\left( T_{\mu\nu}+\langle \hat T_{\mu\nu}\rangle  \right)k^\mu k^\nu\,,
\end{align}
where we have included both the classical and quantum contributions. For classical matter, one expects $\text{NEC}\geq 0$, while it is well known that quantum fields can violate this.

In the star's interior, but far from the inner light ring, the NEC will be positive, since quantum effects there are negligible. In order to investigate the behavior of the NEC in the vicinity of the inner photon sphere as we approach the Buchdahl bound, we choose the null vector as $k^\mu=(1,k^r,0,0)$. We then compute \eqref{NEC} inside the star, in the limit $\epsilon\to 0$, keeping fixed the ratio $r/\ri$. This yields
\begin{align}\label{NECrstar}
\text{NEC}(r) &= \frac{2d}{27\pi^2 G^4R^4} \frac{\ri^2-r^2}{\ri^2+r^2}+\mathcal{O}(\epsilon)\,.
\end{align}

This is effectively ``tracking'' the NEC in the region around the inner photon sphere as the configuration approaches the Buchdahl bound, since $\ri\to 0$ in this limit. The NEC clearly changes sign at the light ring and is thus violated. Notice that the classical contribution is subdominant in this limit and is contained in the subleading orders. On the other hand, choosing $k^\mu$ along the $(t,\phi)$ plane does not lead to a violation. 

The analysis above posits an interesting question. Stable light rings have been recently associated with gravitational instabilities~\cite{Keir:2014oka,Cardoso:2014sna}, which would rule out ultracompact objects\,\cite{Cunha:2017qtt}. However, we have shown here that it is precisely this feature that enhances the quantum effects there, leading to the violation of energy conditions and to significant backreaction. Exploring this interaction at the nonlinear level is an interesting direction.

\section{Comments on Buchdahl's theorem.}

Buchdahl's theorem relies on several assumptions as stated in the introduction. Our results show that QFT in curved spacetime violates two of these assumptions, namely isotropy of the matter and the effective equations of motion.

As we have seen in \eqref{Tmunuq} and is illustrated in Fig. \ref{fig:plot}, the renormalized vacuum stress tensor of the quantum fields is not isotropic, thus violating one of the assumptions of Buchdahl's theorem. Anisotropic versions of Buchdahl's bound exist but they require extra assumptions\,\cite{Guven:1999wm,Ivanov:2002xf,Barraco:2003jq,Boehmer:2006ye,Andreasson:2007ck}. These typically take the form of energy conditions, with the strength of the bound depending on the strength of the conditions. Here, we have shown that quantum fields violate energy conditions in the probe approximation. We leave it for future work to examine whether the equations including backreaction violate the assumptions leading to these generalized theorems. 

Second, close to the compactness bound the relevant equations of motion to solve are \eqref{semi-cl-einstein}, rather than the classical Einstein equations. These differ by the presence of the quantum source which, as we have shown, becomes the dominant term in the Buchdahl limit. This contribution depends explicitly on the curvature tensors, and therefore the differential equations to solve are of a different nature than the purely classical ones. 

This last feature has an alternative description in terms of quadratic gravity. For our specific background, we have shown that among the terms that determine $\langle \hat T_{\mu\nu} \rangle$ in (\ref{Tmumu}) and (\ref{Tmunuq}), only those controlled by $d$ diverge faster than the classical $T_{\mu\nu}$ as $\epsilon\to0$. The ones associated with $a$ diverge with the same power as the classical terms, but they come with a coefficient that is very small for astrophysical objects. Now as anticipated, $d$ is a scheme-dependent parameter that can be generated by adding the counterterm $-\frac{d}{12(4\pi)^2}\mathcal R^2$ to the Lagrangian. This means that our results can also be interpreted as coming from quadratic corrections to Einstein's gravity. The Weyl flatness of the background, then, is not essential to find the leading terms of $\braket{\hat T_{\mu\nu}}$.

This two-faced interpretation is akin to Starobinsky's inflation \cite{Starobinsky:1980te}, initially formulated in terms of the backreaction of quantum fields, then as $\mathcal R^2$ gravity (in the Jordan frame) or Einstein gravity coupled to a scalar field (in the Einstein frame). In the latter picture, the stability of the scalar field requires the condition $d>0$, the same we found and discussed earlier.

It is worth noticing that Buchdahl's theorem holds in a local form as $\frac{r}{Gm(r)}\geq \frac{9}{4}$, where the radius and mass of the star are replaced by an arbitrary coordinate radius $r$ and the Misner-Sharp mass $m(r)=4\pi \int_0^r dr\, r^2\rho$ contained within it, provided the assumptions are met inside that sphere. For example, the star could consist of an incompressible dense core surrounded by an external crust obeying a softer equation of state. Our results also apply to this generalized scenario.

Interesting recent work has also considered quantum fields in the Buchdahl limit\,\cite{Carballo-Rubio:2017tlh,Arrechea:2021xkp,Arrechea:2021pvg} in the approximation of a two-dimensional reduction. This corresponds to the $s$-wave ($\ell=0$) sector, and leaves the stress tensor undetermined up to an arbitrary function. Our results differ from theirs in that \eqref{Tmunuq} fully captures the $(3+1)$-dimensional features, leaving no functional freedom. For other applications of similar techniques see \cite{Mottola:2006ew,Kawai:2017txu,Beltran-Palau:2022nec}.

\section{Summary}

We have investigated the universal behavior of QFT in the interior of very compact stars. A useful arena to probe this is the strong gravity regime close to Buchdahl's limit that, classically, sets an upper bound on the compactness of static, spherically symmetric spheres in General Relativity. As a proxy for this, we have worked with the constant-density Schwarzschild interior solution.

Motivated by the trapping of classical waves in this metric close to Buchdahl's limit, we have studied quantum fields propagating on this background in the approximation of no backreaction. Exploiting the conformal flatness of this solution, we have evaluated the full renormalized stress tensor \eqref{Tmunuq} for Conformal Field Theories. This depends on two coefficients $a$ and $d$, the latter of which is not fixed by the theory in question.

The vacuum renormalized stress tensor \eqref{Tmunuqstar} is not isotropic, since the radial and angular pressures are different. The sign of the energy density is opposite to that of the pressures. Its components acquire their maximum magnitude at the origin, and fall steeply around the inner light ring, as shown in \eqref{rhofalloff}.

As we approach the Buchdahl limit, the $d$ term of the renormalized stress tensor \eqref{rhoq0}-\eqref{pq0} diverges faster than the classical source \eqref{rhoclass0}-\eqref{pclass0}, meaning that quantum fields respond \textit{stronger} to changes in compactness than their classical counterpart. The crossover when classical and quantum contributions are of the same order happens when the curvature radius is $\sim \sqrt{|d|} \ell_P$. The radial Null Energy Condition---including both classical and quantum contributions---changes sign at the inner photon sphere as shown in \eqref{NECrstar}, and is thus violated inside the star. Whether the scales involved are Planckian or not depends on the value of $d$. If $d\ll 1$, then we cannot trust our semi-classical analysis. On the other hand, if $d\gg 1$, the effects of the QFT \textit{cannot} be ignored in this regime. 

We emphasize that the enhancement of quantum effects discussed here happens as the surface of the star approaches the Buchdahl radius $9GM/4$ instead of $2GM$. Moreover, the effect of the quantum fields is localized in a small region around the center---the inner light ring---and not the surface. This is \textit{different} from ultracompact objects close to the Schwarzschild radius. There, the renormalized stress tensor in the Boulware vacuum is well known to diverge at the surface as the star approaches the black hole limit\,\cite{Barcelo:2007yk}.

The isotropy assumption used in Buchdahl's theorem is violated by vacuum quantum fields. Whether the conditions leading to the anisotropic generalizations of this bound hold or not requires further investigation.

We have not attempted to solve the semiclassical equations \eqref{semi-cl-einstein} here. Nevertheless, our results suggests that if $d>0$, quantum fields act by decreasing the curvature, suggesting that a self-consistent solution to these equations might exist that avoids curvature singularities.

It is intriguing to wonder whether quantum physics may play yet another, unexpected, role in the determination of the maximum mass of compact stars. 

\begin{acknowledgments}
\paragraph{Acknowledgments.} We thank Max Ba\~{n}ados, Pablo Bosch, Alejandra Castro, Jan de Boer and Erik Verlinde for insightful discussions. We also thank Daniel Baumann and Vitor Cardoso for feedback on the manuscript. We are particularly grateful to Ben Freivogel for extensive discussions. 
\end{acknowledgments}

\bibliography{QFT_star_bib_2.bib}

\end{document}